\documentclass{iaus}
\usepackage{epsfig}

  \checkfont{eurm10}
  \iffontfound
    \IfFileExists{upmath.sty}
      {\typeout{^^JFound AMS Euler Roman fonts on the system,
                   using the 'upmath' package.^^J}%
       \usepackage{upmath}}
      {\typeout{^^JFound AMS Euler Roman fonts on the system, but you
                   dont seem to have the}%
       \typeout{'upmath' package installed. iaus.cls can take advantage
                 of these fonts,^^Jif you use 'upmath' package.^^J}%
      }
  \else
  \fi

  \checkfont{msam10}
  \iffontfound
    \IfFileExists{amssymb.sty}
      {\typeout{^^JFound AMS Symbol fonts on the system, using the
                'amssymb' package.^^J}%
       \usepackage{amssymb}%

      }{}
  \fi

  \IfFileExists{amsbsy.sty}
    {\typeout{^^JFound the 'amsbsy' package on the system, using it.^^J}%
     \usepackage{amsbsy}}
    {\providecommand\boldsymbol[1]{\mbox{\boldmath $##1$}}}

\newsavebox{\astrutbox}
\sbox{\astrutbox}{\rule[-5pt]{0pt}{20pt}}

\newcommand\etal{\mbox{\textit{et al.}}}

\newcommand\eg{e.g.\ }

\title{Jet-Induced Star Formation:\\
Good News From Big, Bad Black Holes}

\author[Wil van Breugel{\it et al.\/}]%
{Wil van Breugel$^1$,
Chris Fragile$^1$,
Stephen Croft$^1$\break
Wim de Vries$^{1,2}$,
Peter Anninos$^1$,
\and Stephen Murray$^1$}

\affiliation{$^1$University of California, Lawrence Livermore National
Laboratory, L-413, P.O. Box 808, Livermore, CA 94550\\
$^2$University of California at Davis, One Shields Ave, Davis, CA
95616\\
USA email:
wil@igpp.ucllnl.org,fragile1@llnl.gov,scroft@igpp.ucllnl.org,\break 
wdevries@igpp.ucllnl.org,anninos1@llnl.gov,murray8@llnl.gov}

\pubyear{2004}
\volume{222}
\pagerange{1--8}
\date{?? and in revised form ??}
\jname{The Interplay among Black Holes, Stars and ISM \\in Galactic Nuclei}
\editors{Th. Storchi Bergmann, L.C. Ho \& H.R. Schmitt, eds.}
\begin{document}

\maketitle

\begin{abstract}
We discuss obbservations and 
numerical simulations which show that radiative shocks in
jet-cloud collisions can trigger the collapse of intergalactic clouds and
subsequent star formation in low luminosity, 'FR-I' type, radio galaxies. 
\end{abstract}

\section{Introduction}

Many, if not all, galaxies have massive - supermassive ($10^6 - 10^9$
Solar mass) black holes at their centers.  These black holes become
`active' sources of energy - Active Galactic Nuclei (AGN) - when they
accrete material from their surroundings in hot accretion disks. These
accretion disks emit high energy radiation and particles, drive high
velocity winds and can form collimated, relativistic, jets.  Heating,
radiative ablation, and shocks will affect the ambient interstellar
medium (ISM) or even intergalactic medium (IGM) out to very large
distances. High pressure starbursts (100 - 1000 times our Galaxy)
produce large numbers of supernovae which
affect their surrounding ISM in similar ways.

This high energy feedback from active black holes and exploding stars
may have a profoud effect on the ISM in their parent galaxies.  In fact,
it has been proposed (Silk and Rees 1998) that this might
self-regulate galaxy and black hole growth and explain
the observed close correlation between the masses of supermassive black
holes and the stellar bulges of their parent galaxies (\eg\ Magorrian
\etal\ 1998).  Studies of high energy feedback on interstellar and
intergalactic media are therefore an important part of observational and
theoretical work on the formation and evolution of galaxies and their
central black holes.

Most of this work focuses on 'negative' feedback, where intense radiation
fields and starburst winds limit furher star formation and galaxy/black
hole growth. But AGN may also provide 'positive' feedback. In particular,
jets and lobes from radio sources may drive radiative shocks into
dense clouds triggering {\it extra} star formation.

Radio jets can trigger star formation in two different ways. Although the
powerful, collimated jets in 'FR-II' type (Fanaroff and Riley 1974) radio galaxies 
punch through clouds in the ISM/IGM, their
slowly expanding radio lobes can trigger star formation in cocoons around
them (Bicknell \etal\ 2000).  On the other hand the less powerful jets
in 'FR-I' type radio galaxies 
may trigger star formation even in head-on cloud collisions (Fig. 1;
van Breugel \etal\ 1985; Croft \etal\ 2004).  In general, in the early
Universe, when galaxies were still forming and gas densities were much
higher, jet-induced star formation may have been relatively common. Evidence
for that is strongly suggested by the discovery of the radio/optical
alignment effect in very distant radio galaxies (McCarthy \etal\ 1987; Chambers et al 1987).

\section{Jet-induced star formation in nearby radio galaxies}

\begin{figure}
{\center
\psfig{file=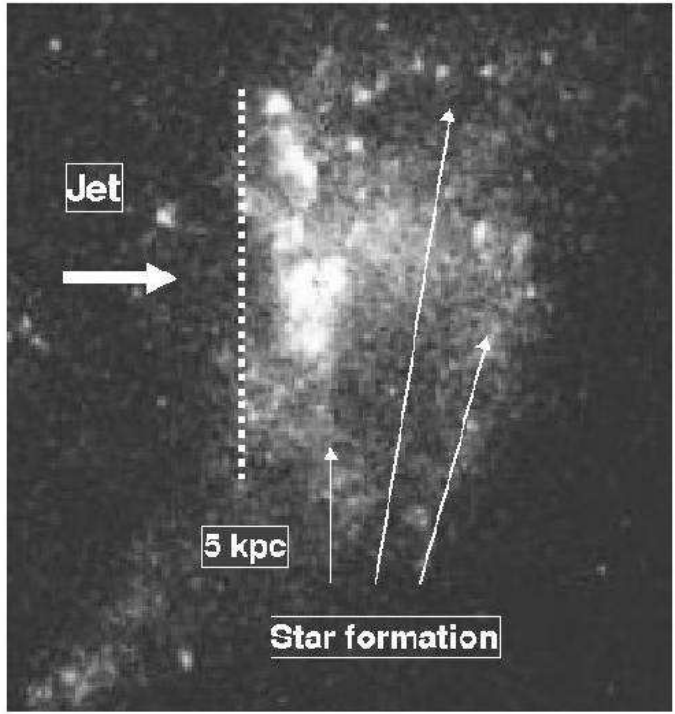,width=10cm}
\caption{HST observation of induced star formation at end of a
jet (Croft \etal\ 2004)}
}
\end{figure}

The first evidence for jet-induced star formation radio galaxies was
found in the nearest, FR-I type, radio galaxy Centaurus A.
Recent observations with the Hubble Space Telescope
have confirmed that there are about half a dozen young ($< 15$ Myr) OB
associations near filaments of ionized gas located between the radio
jet and a large HI cloud (Mould et al. 2000). Star forming regions
associated with radio sources have also been found in cooling flow
clusters, with the best example being in Abell~1795 (McNamara 2002).

One of the most spectacular observed jet-induced starbursts is
'Minkowski's Object' associated with the elliptical galaxy NGC~541 in
the
 cluster of galaxies Abell~194 (Fig. 1). 
Its morphology
 is strongly suggestive of a
collision between the FR-I type jet from
 NGC~541 and a dense cloud:
M.O. has the same overall diameter as the
 jet, appears wrapped around
the end of the jet, has bright emission in
 the upstream direction and
filamentary structure down-stream where the
 jet appears disrupted.
Spectroscopically M.O. looks like an HII region,
 resembling starburst
galaxies. The H$\alpha$ luminosity suggests a modest
 star formation rate of
0.3 M/yr. VLA observations (van Breugel and van
 Gorkom, unpublished)
show two detections down-stream from the jet-cloud
 collision site,
indicating a total HI mass of $\sim 3 \times 10^8$
 M$_\odot$.

\section{Numerical simulations of jet-cloud collisions}

\begin{figure}
{\center
{\psfig{file=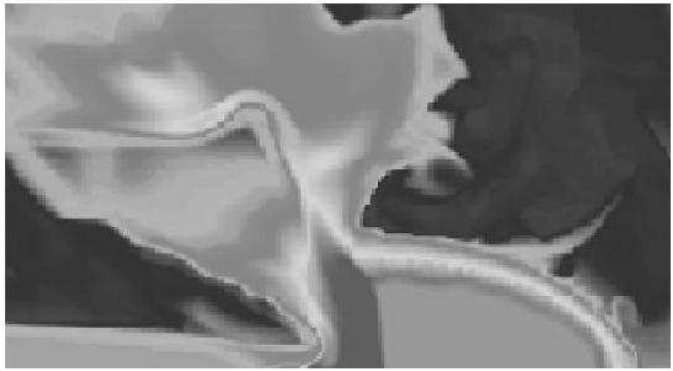,width=10cm}}
\caption{Numerical simulation of jet / cloud collision using
COSMOS (Fragile et al 2004). Radiative shocks trigger star formation at
front (left) and the sides}
}
\end{figure}

Numerical simulations are essential if we want to learn more about
the physical conditions in radio 
 jets and their environment (density,
temperature,
 composition etc.).  For example, we don't even know what
is {\it inside}
 these jets (e$^-$e$^+$ ?; e$^-$p$^+$ ?; Hujeirat 2004),
and direct
 knowledge about the physical conditions in forming galaxies
will always
 be limited because of their faintness. Previous numerical
simulations
 by several other groups 
did not include effects of
 radiative cooling,
magnetic fields etc.
Radiative cooling is important because it allows compressed clouds to
collapse and break up in numerous, dense cold fragments which survive
for many timescales and are presumably the precursors to star formation
(Fragile et al 2004; Mellema et al 2002).
Previous
work has also focused primarily on FR-II type radio galaxies even though,
because of their proximity, much better observational data can be
obtained for FR-I type jet-induced starbursts such as M.O. 

Fragile et
al (2004) used the LLNL developed, multi-dimensional, multi-physics,
massively parallel 'COSMOS' numerical simulations package 
to investigate both FR-I and FR-II
type jet-induced star formation systems.
To simulate the jet induced star formation in M.O. Fragile et al (2004)
assumed that NGC~541 is surrounded by a multi-phase medium resembling
cluster atmospheres (e.g. Ferland et al. 2002).  Specifically, it was
assumed that the FR-I jet interacts with a moderate density,
hot `mother'-cloud with a semimajor axis of 10 kpc, a semiminor axis of
5 kpc, a density $n_{mcl} = 0.1$ cm$^{-3}$, and temperature $T=10^6$ K.
This corresponds to an initial total cloud mass of $\approx 10^9
M_\odot$.
Within this mother-cloud denser, warm clouds were assumed to be embedded
with typical sizes of 100 pc, $n_{cl} = 10$ cm$^{-3}$, and temperature
$T=10^4$ K.
The detailed radio and X-ray study of the proto-typical FRI-type radio
galaxy 3C31 by Laing and Bridle (2002) was used to estimate plausible
jet
parameters near M.O., at $\sim 15$ kpc from the NGC~541 AGN.

The collision with the mother-cloud triggers a nearly planar shock down
the long axis of the cloud (Fig. 2).  As the bow
shock from the jet wraps around, it also triggers shocks along the
sides of the cloud. This may explain the filamentary nature of the
star-forming region seen in M.O. downstream from the jet (Fig. 1).
To determine whether the clouds embedded within the mother-cloud
would indeed collapse and form stars we ran a number of simulations to
explore the density / velocity parameter space. We found that for shock
velocities between $v_{sh}$  = 1,000 - 10,000 km s$^{-1}$ and $n_{cl} =
1 - 100 $ cm$^{-3}$ the cooling times were sufficiently short that star
formation could occur.

\section{Conclusions}

Our results show that M.O. could be due to star formation which was
triggered by the collapse of dense Intergalactic Medium (IGM) gas clouds
due to a collision with the jet. Star formation in the IGM near expanding
radio lobes has been seen in `cooling' flow clusters (McNamara 2002).
Although A~194 is a not a cooling flow cluster, extended X-ray emission
has been detected with ROSAT and extends along a prominent stellar
bridge which connects three large ellipticals: NGC~541 and its double
companion galaxy  NGC~545/547 associated with 3C40 to the North-East
(Lazzati \etal\ 1998; Nikogossyan \etal\ 1999). M.O. appears to be
located in this bridge, where regions of overdense gas would not be
unexpected. Further observations to search for evidence of overdense gas
in this stellar bridge are underway.

\begin{acknowledgments}
This work was performed under the auspices of the U.S. Department of
Energy, National Nuclear Security Administration by the University
of California, Lawrence Livermore National Laboratory under contract
No. W-7405-Eng-48.  W.v.B.\ also acknowledges NASA grants GO~9779 and
GO3-4150X in support of high-redshift radio galaxy research with HST
and Chandra.
\end{acknowledgments}


\medskip

\noindent{\bf References}

\noindent
Bicknell, G. et al. 2000, ApJ, 540, 678 \\
Chambers, K. et al. 1987, Nature, 329, 604 \\
Croft, S. et al. 2004, in prep \\
Fanaroff, B. L. \& Riley, J. M. 1974, MNRAS, 167, 31P \\
Ferland, G. J. et al. 2002, MNRAS, 333, 876 \\
Fragile, P. C. et al. 2004, ApJ, 604, 74 \\
Hujeirat, A. 2004, A\&A, 416, 423 \\
Laing, R. A. \& Bridle, A. H. 2002, MNRAS, 336, 1161 \\
Lazzati, D. et al 1998, A\&A, 339, 52 \\
Mould, J. R. et al. 2000, ApJ, 536, 266 \\
Magorrian, J. 1998, AJ, 115, 2285 \\
McCarthy, P. J. et al. 1987, ApJ, 321, L29 \\
McNamara, B. R. 2002, New Astronomy Review, 46, 141 \\
Mellema, G. et al. 2002, A\&A, 395, L13 \\
Nicogossyan, E. et al 1999, A\&A, 349 97 \\
Silk, J. \& Rees, M. 1998, A\&A, 331, L1 \\
van Breugel, W. et al. 1985, ApJ293, 83 \\

\end{document}